\documentclass[preprint,3p,onecolumn]{elsarticle}
\usepackage{amssymb}
\usepackage{amsmath}
\usepackage[linesnumbered,ruled,vlined]{algorithm2e}
\usepackage{graphicx}
\usepackage{setspace}
\usepackage{caption}
\usepackage{array}
\usepackage[pagewise]{lineno}
\usepackage{subcaption}
\newcolumntype{C}[1]{>{\centering}m{#1}}

\begin{document}

\begin{frontmatter}

\title{Modelling and Performance Evaluation of Stealthy False Data Injection Attacks on Smart Grid in the Presence of Corrupted Measurements}
\author[rvt]{Adnan Anwar\corref{cor1}}
\ead{Adnan.Anwar@adfa.edu.au}
\author[rvt]{Abdun Naser Mahmood}
\author[rvt]{Mark Pickering}
\address[rvt]{School of Engineering and Information Technology (SEIT), \\The University of New South Wales Australia, Canberra, ACT 2610, Australia}

\cortext[cor1]{Corresponding author. Tel: {+61451001357}}

\begin{abstract}
The false data injection (FDI) attack cannot be detected by the traditional anomaly detection techniques used in the energy system state estimators. In this paper, we demonstrate how FDI attacks can be constructed blindly, i.e., without system knowledge; including topological connectivity and line reactance information. Our analysis reveal that existing FDI attacks become detectable (consequently unsuccessful) by the state estimator if the data contains grossly corrupted measurements such as device malfunction and communication errors. The proposed sparse optimization based stealthy attacks construction strategy overcomes this limitation by separating the gross errors from the measurement matrix. Extensive theoretical modelling and experimental evaluation show that the proposed technique performs more stealthily (has less relative error) and efficiently (fast enough to maintain time requirement) compared to other methods on IEEE benchmark test systems. \end{abstract}
\begin{keyword}
Smart grid, false data injection, blind attack, principal component analysis (PCA).
\end{keyword}

\end{frontmatter}

\section{Introduction}
Recently, smart grid cyber-security has come to the forefront of national security priorities. Several power system anomalies have been attributed to cyber-attacks, highlighting the importance of research on the impact of new kinds of attacks on complex power systems. Nation states and utilities are increasingly concerned about power system integrity, privacy and confidentiality, particularly in the aftermath of the infamous `Stuxnet worm'~\cite{stuxnet} attack in 2010. Recently, the ‘Industrial Control Systems Cyber Emergency Response Team (ICS-CERT)’ reported that among the $245$ cyber incidents across all sectors of the critical infrastructure in the fiscal year 2014, the majority (32\% or 79 incidents) were in the energy sector~\cite{ICSreport2015}. In today's smart grid, the physical energy system and the information and communications technology based cyber system are highly coupled which introduces new security threats.

The state estimator (SE) is a key operational module used in a smart grid energy management system (EMS) to estimate the power system states (e.g., voltage magnitudes and angles) from sensor measurements by minimizing estimation error. A bad data detector module (BDD) works in conjunction with the SE module to identify any anomalies in the measurement data~\cite{abur2004power}. Recent studies have revealed that these critical operational modules (e.g., SE and BDD) are vulnerable to a class of cyber-attack~\cite{Liu:2009:FDI:1653662.1653666,yu7001709,AnwarBookApelVi}, known as a \textit{false data injection} (FDI) attack. In a seminal work~\cite{Liu:2009:FDI:1653662.1653666}, Liu et al. have shown that an attacker can construct a stealthy FDI attack that cannot be detected by traditional anomaly detection modules (BDD) of an EMS. Due to this hidden FDI attack, the SE estimates wrong system states, which misleads the system operator in taking wrong operational decisions that can lead to degradation of the system efficiency, reliability and may trigger cascading failures. Hence, significant research has been carried out on FDI attacks to investigate their stealthiness and construction strategies~\cite{Liu:2011:FDI:1952982.1952995,Kosut5622045,Hug6275516,Kim6840319,Qingyu6490324,Ozay6547838,Esmalifalak6102326,Kim6996007,yu7001709,Anwar2016pes}, along with prospective detection and prevention measures~\cite{Suzhi6787030,Qingyu6490324,Hug6275516,Jokar6655271,Pan7063234}.
The objective of this work is to demonstrate circumstances when attacks no longer remain stealthy, and a potential strategy that an attacker can apply to circumvent it. Next, we provide more background on this problem followed by a discussion on the significance and novelty of the proposed approach.

\subsection{Related Work}

There is a growing number of research papers on different aspects of FDI attacks. While some (e.g,~\cite{Liu:2011:FDI:1952982.1952995,yu7001709}) consider that all measurement devices are vulnerable to cyber attacks, others assume that a subset of measurement devices are compromised~\cite{Bi6787030}. Some attack detection strategies, e.g., Liu et al. assume that phasor measurement unit (PMU) data is not compromised and build their model around this assumption~\cite{Liu6740901,Liu6655269}. Examples of PMU attack include false data injection, and GPS based spoofing attack. For example, intelligent cyber-attackers can gain access through networked devices and gain the access to PMU by false-using of IP Multicast routing protocols~\cite{Mousavian6816087}.
In addition, authors describe in detail~\cite{Shepard2012146} how a GPS based PMU can be attacked and wrong data can be injected.
In~\cite{Shepard2012146}, it was reported that currently there are no defences available against these types of PMU attacks.
Apart from the PMUs, supervisory control and data acquisition (SCADA) protocols (e.g., modbus), that is widely used to collect remote sensor data, are also vulnerable to cyber attacks~\cite{Queiroz6009221}.

There are mainly two classes of FDI attacks: one class of attacks make the assumption that various degrees of the system knowledge is known to the attacker, (e.g., network topology, branch and node information, etc.) that can be used for attack construction~\cite{Liu:2011:FDI:1952982.1952995,Kosut5622045,Hug6275516,Kim6840319,Qingyu6490324,Ozay6547838}; another class of FDI attack relaxes the requirement of prerequisite knowledge and these attacks are more realistic~\cite{yu7001709,Anwar2016PAISI}.
For example, in~\cite{Liu:2011:FDI:1952982.1952995}, it is discussed how an FDI attack can be constructed using power system information which includes power system topology and line parameters, known as the \textit{system Jacobian}. Based on the system Jacobian, an attacker can introduce arbitrary errors in the calculation of the state variables by the state estimator module, through injection of false information into sensor measurement data. In~\cite{Kosut5622045}, an efficient algorithm using a graph theoretic approach was proposed that can construct an stealthy attack. Another graph based attack strategy utilizing power system topology was proposed in~\cite{Hug6275516}. A \textit{data framing} FDI attack was proposed in~\cite{Kim6840319}, where it was shown that an attacker can inject data in such a way that the EMS identifies correctly functioning measurement devices as a false data injection source. In our previous work~\cite{Anwar2015jrnl1}, we have described a relationship between the power system stability indices and the FDI attacks in a smart grid environment when the attacks are injected through smart meters. In~\cite{Qingyu6490324}, an attack was defined that maximizes the deviations of the system states. In~\cite{Ozay6547838}, a~\textit{collective sparse attack} strategy was proposed where state variables in the same cluster are attacked by the same attack vector. However, in a real-world system, it is very difficult to obtain the detailed system information required by these types of FDI attacks. To overcome this limitation, authors in~\cite{Rahman6503599} assume that the attacker has partial or limited information about the system topology and power system parameters. In~\cite{Esmalifalak6102326}, a method based on independent component analysis (ICA) was proposed to construct a stealthy attack with low detection rate. In that work, it was assumed that the attacker has no knowledge of the system information and a stealthy attack was constructed based on the measurement matrix. A similar method for attack construction considering both full measurements and partial measurements was proposed in~\cite{Kim6996007}. Another data-driven attack that did not need any prior system Jacobian information, is studied in~\cite{yu7001709}.
In~\cite{Kim6996007,yu7001709}, the authors show that a power system knowledge-free FDI attack can be constructed using a subspace method which exhibits stealthiness. Most recently in~\cite{yu7001709}, the authors proposed a Principal Component Analysis (PCA) based blind FDI attack construction strategy that was shown to successfully and stealthily attack the system by the measurements using its subspace information. There are two main drawbacks of this approach. First, the key parameter of the PCA based approach requires the proper selection of the dimensionality parameter, which requires the knowledge of the total number of states of any system (\textit{rank} of Jacobian matrix). Since the total number of states of a system is unknown to an outside attacker, this strategy cannot be used as a zero knowledge based attacks. Second, these data-driven attack strategies assume that the measurements are noiseless or include random noise that follows Gaussian distribution. However, like other application areas (e.g., image processing, bio-informatics), gross errors (e.g., errors that do not follow Gaussian distribution) are also ubiquitous in smart grid measurement data due to sensor failures or communication errors.
Theoretically, it has been proven~\cite{Candes:2011:RPC:1970392.1970395} that the accuracy of PCA is affected if the data contains gross errors. Furthermore, in Section 6.3, we demonstrate that PCA based attack strategies are ineffective in the presence of gross errors.

\subsection{Contribution}
In the context of stealthy attack generation without system knowledge (e.g., topology, system states), this paper has the following contributions to the state-of-the-art research:

(1)	Existing attack strategies (see section 3) require the system Jacobian matrix \textbf{H}, which represents the interconnectivity information (as well as resistance/reactance values) of all the buses and branches of a power grid. Recent techniques have emerged, where an adversary can construct a blind attack without the system Jacobian information \textbf{H}. However, as discussed in the \textit{first} drawback of existing literature in Section 1.1, these techniques~\cite{Kim6996007,yu7001709} assume that the number of system states\textemdash that is the rank of \textbf{H}~\cite{yu7001709} \textemdash is known to the adversary. In practice, neither the information \textbf{H} nor the number of system states ($n$) is available to the adversary.  This paper proposes a technique that can create an attack vector without knowing \textbf{H} or $n$.

(2) Existing FDI attack techniques assume that the measurement data contains only Gaussian noise. However, as discussed in the \textit{second} drawback of existing literature in Section 1.1, previous research did not consider device malfunction or missing data, which we refer to \textit{gross} error. With extensive experiments, we demonstrate in Section 6.3 that existing data-driven FDI attacks do not remain stealthy in the presence of gross errors.

(3) This paper presents a \textit{blind} FDI attack construction strategy in the presence of gross errors and Gaussian noises using sparse optimization technique. We show that the proposed attack construction technique using augmented lagrange multiplier (ALM) method can generate stealthy FDI attacks successfully in the presence of grossly corrupted measurements.

(4) We also evaluate the performance of the ALM based technique with three other effective sparse optimization techniques: accelerated proximal gradient (APG)~\cite{Lin2009fast}, singular value thresholding (SVT)~\cite{Cai:2010:SVT:1898437.1898451} and the dual method~\cite{Lin2009fast}. Extensive experimental evaluation on benchmark IEEE test systems, the ALM based attack construction technique shows its effectiveness over other exixting methods in terms of both accuracy and efficiency.

The organization of this paper is as follows- in Section~\ref{sgmm}, a smart grid measurement model is discussed which provides the background on the SE and BDD module. In Section 3, an adversary model is presented, where we first review the attack strategies that need power system \textit{Jacobian} knowledge.
Data-driven blind attack strategy is discussed in Section 4. The proposed attack strategy in the presence of grossly corrupted measurements is presented in Section 5. Extensive experimental results considering benchmark test systems and multiple scenarios are presented in Section~\ref{RnD}. The paper concludes with some brief remarks in Section~\ref{SectionFinish}. The terms and their explanations used in the paper are listed in Table~\ref{Listsymlbl}.
\begin{table}
\caption{Glossary of Terms}\label{Listsymlbl}
\centering
\begin{center}
{\setlength{\extrarowheight}{3pt}%
\begin{tabular}{| c  | p{6.2cm}  || c  | p{5.6cm}  | }
\hline
Notation 	& Description & Notation 	& Description \\ \hline
$\mathbf{z}$	& Vector of original measurement& $\mathbf{z'}$	& Vector of compromised measurement\\ \hline
$\mathbf{x}$	& vector of original system states& $\mathbf{x'}$	& vector of compromised system states\\ \hline
$\mathbf{e}$	& vector of gaussian noises & $\mathbf{u}$	& vector of principal components\\ \hline
$\mathbf{a}_{pca}$	& attack vector & $\mathbf{r}$	& residual vector\\ \hline
${\psi}$	& degree of freedom & $\tau$	& bad data detection threshold\\ \hline
$m$	& total number of measurement devices & $n$	& total number of system states\\ \hline
${\sigma_i}$	& standard deviation of i$^{th}$ measurement & $\lambda$	& sparsity regularization parameter\\ \hline
$\mathbf{H}_{pca}$	& reduced PCA transformation matrix & $\mathbf{\tilde{M}}$	& PCA transformation matrix\\ \hline
$\mathbf{{M}}$	& measurement matrix with gross errors and gaussian noises & $\mathbf{{A}}$	& low-rank measurement matrix without gross errors\\ \hline
$\mathbf{{E}}$	& sparse gross error matrix & $\mathbf{H}$	& system Jacobian matrix\\ \hline
$\boldsymbol{{\gamma}}$	& Lagrange multiplier & $\boldsymbol{\mu}$	& dual variable\\ \hline
$\mathbf{{U,V}}$	& unitary matrix of SVD operation & $\mathbf{S}$	& diagonal matrix of SVD operation\\ \hline
${\sum\nolimits}$	& covariance matrix & $\mathbf{I}$	& Identity matrix\\ \hline
$\xi$	& soft-thresholding operator & P$_{mis}$	& probability of false negative\\ \hline
$\mathcal{C}$(\textbf{X})	& column space of \textbf{X} & $\mathbb{E}$(\textbf{X})	& expected value of \textbf{X}\\ \hline \hline
SE	& state estimation &  FDI	& false data injection\\ \hline
ALM	& augmented Lagrange multiplier &  PCA	& principal component analysis\\ \hline
ICS	& industrial control system &  CERT	& cyber emergency response team\\ \hline
EMS	& energy management system &  BDD	& bad data detectot\\ \hline
PMU	& phasor measurement unit  &  GPS	& global positioning system\\ \hline
SCADA	& supervisory control and data acquisition  &  WSSE	& weighted sum of squared error\\ \hline
ICA	& independent component analysis  &  APG	& accelerated proximal gradient \\ \hline
DC	& direct current  &  SNR	& signal-to-noise ratio\\ \hline
IP	& internet protocol  &  LNR	& largest normalized residual\\ \hline
SVT	& singular value thresholding  &  SVD	& singular value decomposition\\ \hline
\end{tabular}}
\end{center}
\end{table}

\section{Preliminaries on Smart Grid Measurement Model}\label{sgmm}
The real-time operation of an energy management system (EMS) depends on the measurement data obtained from a supervisory control and data acquisition (SCADA) system. The operational functionalities (optimal power flow, economic dispatch, contingency analysis, etc) of a smart grid require knowledge of the power system states (typically, voltage magnitudes and angles) for making real-time operational decisions. However, power system measurement signals are often noisy and subject to missing values. For reliable power system operation, treatment of measurement data is necessary to obtain system states by removing noises and anomalous data. In an EMS, the state estimator (SE) and bad data detector (BDD) are responsible for performing these tasks. In the presence of Gaussian noise (which is widely used in the literature as these types of errors are common in measurement data), the measurement vector \textbf{{z}} can be represented as:
\begin{equation}\label{}
\bf{{z}}=h(\bf{{x}})+\bf{{e}}
\end{equation}
where \textbf{h($\cdot$)} is the system \textit{Jacobian} matrix which defines the non-linear relationship of the measurements and the system states (\textbf{x}). Here, \textbf{e} is a vector of zero mean Gaussian noise elements. Generally, it is assumed that the noise elements are independent (so, $\sum_{\textbf{e}}$=\textbf{R}=\textbf{$diag\{\sigma_1^2,\sigma_2^2,...,\sigma_m^2\}$)}~\cite{abur2004power}, where $\sigma_i$ is the standard deviation of each measurement $i$. Generally, Direct Current (DC) power flow model is widely used by the power engineers as well as Smart Grid cyber-security researchers~\cite{yu7001709,Kim6996007,Liu6740901} to describe the linear approximation of Alternative Current (AC)~\cite{abur2004power} power flow model. The DC approximation is widely accepted as a substitute for AC model because (i) it guarantees faster convergence; (ii) reduces algorithmic complexities of power flow analysis, (iii) particularly used for transmission system analysis as it produces highly accurate results~\cite{abur2004power,Liu:2011:FDI:1952982.1952995}. Considering a DC approximation, the linearized measurement model becomes as follows~\cite{abur2004power,yu7001709}:
\begin{equation}\label{dcselbl}
\bf{{z}}=H{\bf{x}}+\bf{{e}}
\end{equation}
where \textbf{H} is the system \textit{Jacobian} matrix. For $m$ number of measurement devices and $n$ system states, the dimension of \textbf{H} matrix is $m \times n$. A system will be \textit{observable} if \textbf{H} is a full rank matrix which leads to the assumptions that  $m \geq n$ and $rank$(\textbf{H})=$n$~\cite{Kim6996007}.

\subsection{The State Estimator}
In the state estimator module, the system state vector \textbf{{x}} is obtained using a \textit{weighted least square (WLS)} estimator which can be formulated as follows:
\begin{equation}\label{wlseq}
\operatornamewithlimits{argmin}_\textbf{x} J(\mathbf{x}) = \frac{1}{\sigma^2}\|\bf{z}-\bf{Hx}\|^2_2 \\
\end{equation}
where the difference between \textbf{z} (measurement data) and \textbf{Hx} (\textbf{H} is the system Jacobian and \textbf{x} is the system states) is called the \textit{residual} \textbf{r}~\cite{abur2004power}.
\begin{equation}\label{}
\bf{r}=\bf{z}-\bf{H}\bf{{x}}
\end{equation}
The problem formulated in (\ref{wlseq}) can be solved iteratively (e.g., using gradient based Newton's method)~\cite{abur2004power}.
\subsection{The Bad Data Detector}
In a power system BDD, the chi-square ($\chi^2$) test is widely used to detect bad measurement data~\cite{abur2004power,Kekatos6340375}. As it is assumed that noise samples follow a \textit{normal} distribution with zero mean and they are independent, ${J(\textbf{x})}$ will follow a ($\chi^2$) distribution with a $\psi$ degree of freedom, where $(\psi=m-n)$~\cite{abur2004power}. Considering a desired significance level (e.g., 95\%), a threshold- $\chi^2_{(m-n),p}$ from the chi-square distribution can be obtained. If there exists an anomalous (bad) data, the value of ${J(\textbf{x})} \geq \chi^2_{(m-n),p}$.
Once the existence of anomalous data is determined, then the largest normalized residual (LNR) test is employed to identify the corrupted data~\cite{abur2004power,Teixeira5717318}:
\begin{equation}
\left\{ \begin{array}{ll}
 ~~$bad data exists,$ & ~~~\textrm{if $\max$~($\frac{1}{\sigma^2}|\textbf{r}^{(i)}|)  >  \tau $}\\
 ~~$no bad data,$ & ~~~\textrm{otherwise}\
\end{array} \right.
\label{anomalydetect}
\end{equation}
where $i\in R^m$ and $\tau$ is the bad data detection threshold~\cite{abur2004power}. In the existing BDD module, an `alarm' is raised if any bad data is detected. Next, the bad data is removed from the measurement vector and the state estimator re-computes the states followed by a BDD operation. This task is continued until all the bad data have been removed.

\section{The Adversary Model with known System Jacobian}
Adversary models for FDI attacks can be classified into two broad categories- (i) a model that requires knowledge of system parameters and topology (system Jacobian), (ii) a model that does not require any information about the system Jacobian. In this section we review the prior model which was proposed by the Liu et al.~\cite{Liu:2009:FDI:1653662.1653666}. According to that model, an attacker with system knowledge $\textbf{H}_{m \times n}$ can strategically inject an attack vector $\textbf{a}_{m\times 1}$ with the measurement signals $\textbf{z}_{m\times 1}$ which cannot be identified by the traditional SE and BDD modules~\cite{Liu:2009:FDI:1653662.1653666}. Suppose, for an attack vector $\textbf{a}_{m\times 1}$, the new corrupted measurements become $\textbf{z}'_{m\times 1}$ =$\textbf{z}_{m\times 1}$  +$\textbf{a}_{m\times 1}$. Based on this manipulated measurements, the SE module will produce wrong system states $\textbf{x}'_{m\times 1}$ instead of the original states $\textbf{x}_{m\times 1}$. The mismatch of system states is considered as ${\textbf{c}}$, where ${\textbf{x}'}$= ${\textbf{x}}$ + ${\textbf{c}}$. Liu et al. shows both theoretically and experimentally that the attack remains hidden if the attack vector satisfy the condition $\textbf{a}$=$\textbf{Hc}$~\cite{Liu:2009:FDI:1653662.1653666}. Following this strategy, the residual of the estimation becomes as below:
\begin{equation}\label{}
\begin{aligned}
& ~~~~~~\|\textbf{z}'-\textbf{Hx}'\| = \|\textbf{z+a}-\textbf{H(x+c)}\| \nonumber  \\
&\Longrightarrow \|\textbf{z}'-\textbf{Hx}'\|  = \|\textbf{z+a}-\textbf{Hx}-\textbf{Hc}\| \nonumber  \\
&\Longrightarrow \|\textbf{z}'-\textbf{Hx}'\|  = \|\textbf{z}-\textbf{Hx}\| ~~~~~(as,~ \textbf{a}=\textbf{Hc}) \\
&\Longrightarrow \textbf{r}_{normal}=\textbf{r}_{attack}
\end{aligned}
\end{equation}
Here, the residual of the attack measurements ($\textbf{r}_{attack}$) is the same as the one without any attack ($\textbf{r}_{normal}$). Hence, the BDD module will fail to detect it using the current statistical testing used in the utilities. Therefore, attack remains hidden which will change the system states affecting critical operational failures.

\section{The Adversary Model with unknown System Jacobian- Blind Attack Strategy}\label{blindSeclbl}

\subsection{Stealthy Attack using Measurements Only}\label{Hunknown}
\begin{figure}[h]
  \centering
    \includegraphics[width=0.9\textwidth]{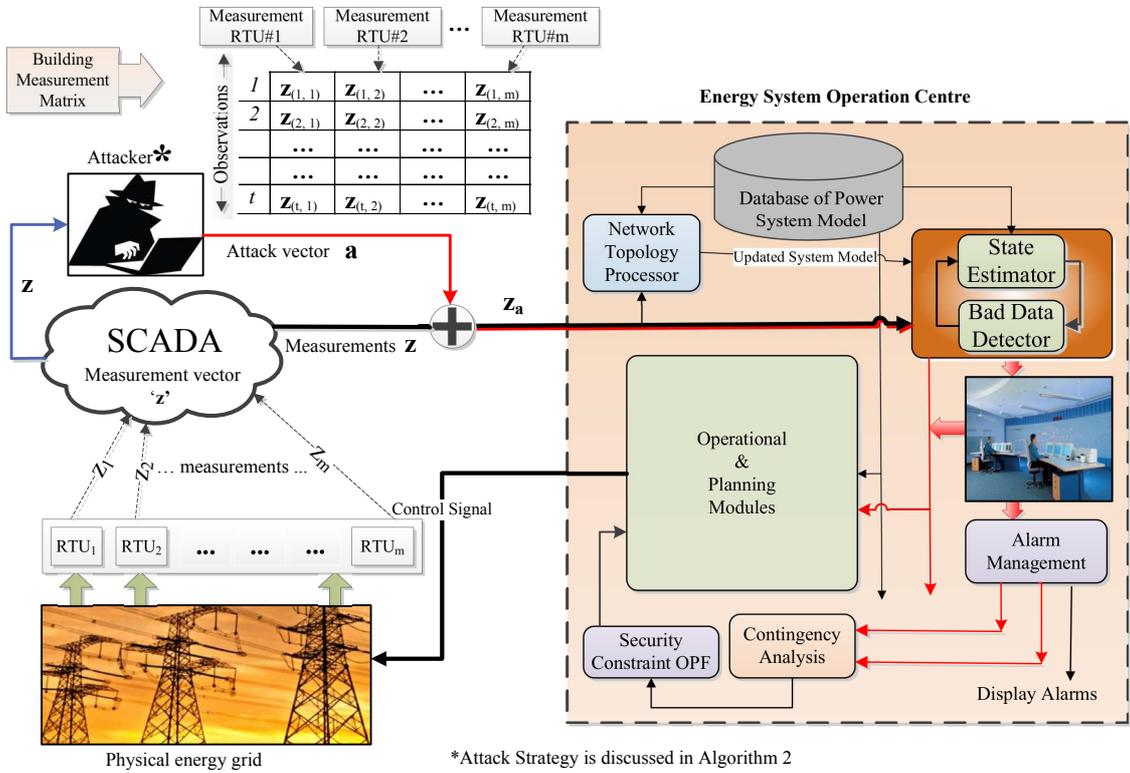}\\
  \caption{Overall architecture of the measurement signal based data-driven FDI attacks}\label{fdialllbl}
\end{figure}

The blind FDI attack strategy is a data-driven approach where the stealthy attack-vector is prepared solely from the measurement matrix. For any blind attack strategy that requires measurement data, the assumption is that all~\cite{yu7001709} or partial (by satisfying observability~\cite{Kim6996007}) measurement data is available to the attacker. An attacker can obtain the data by gaining access to the PMUs (as mentioned earlier in the literature review~\cite{Mousavian6816087}), using man-in-the-middle attack on data in transit (e.g., attacking the router, or by spoofing as fake measurement devices, etc.)~\cite{Wang20131344}. Hence, measurement data \textbf{Z} is known to the adversary.  Measurement signal based blind attack is possible because for a small range of time, the power system consumptions (loads) vary in a small dynamic range~\cite{Esmalifalak6102326}. Unless there is a reconfiguration of the system, power system topology remains the same. Hence, due to the slow dynamic nature for a short period of time, the equivalent knowledge of the topology can be revealed using the correlations among multiple power flow measurements~\cite{Esmalifalak6102326,yu7001709,Kim6996007}. Hence, the attacker monitors the measurement data vector $z=[z_1,z_2,…,z_m]$, where  $z_i$ is the measurement of i-$th$ device. Observing $t$ number of measurement vectors, the attacker constructs the measurement matrix \textbf{Z}$_{t\times m}$, where each row represents an observation of all measurement devices at time \textit{t} (see Fig.~\ref{fdialllbl}). Next, this measurement matrix is used to construct blind attack as discussed below. Fig.~\ref{fdialllbl} shows the conceptual diagram of how an attacker would construct the attack from measurement data \textbf{Z}. The detailed technical procedure of attack construction strategy is also discussed in Algorithm 2.

Recently, Kim et al. in~\cite{Kim6996007} shows that the subspace estimation method can be used successfully to learn the system operating subspace without the need of system knowledge (e.g., topology and system parameters) to generate FDI attacks.
In another recent study, Yu et al. showed that PCA can be used to transform the measurement data into a new subspace, preserving the spatial characteristics as much as possible~\cite{yu7001709}. Using these approaches it is possible to construct FDI attacks based on the data of the new projected space.
If the column space of any matrix is represented as $\mathcal{C}(.)$, designing a stealthy attack is equivalent to finding a nonzero vector in $\mathcal{C}(\textbf{{H}})$~\cite{Kim6996007}. Therefore, kim et al. has pointed out that attack can be constructed using a basis matrix of $\mathcal{C}(\textbf{{H}})$ without knowing $\textbf{H}$~\cite{Kim6996007}. Detailed procedure is explained below:

Consider the time-series measurement matrix $\textbf{Z}_{t\times m}$, where each row represents a time instant (observations) and each column corresponds to the measurement variables, the state vectors $\textbf{x}_1, . . . , \textbf{x}_d$ are independent and identically distributed (i.i.d.), and the noise vectors and the state vectors are uncorrelated. Then, the covariance ($\Sigma_\textbf{{Z}}$) of the measurement matrix becomes as follows~\cite{Kim6996007}:
\begin{equation}\label{dcselbl}
\Sigma_\textbf{{Z}} \triangleq \mathbb{E}[(\textbf{{Z}}-\mathbb{E}[\textbf{{Z}}])(\textbf{{Z}}-\mathbb{E}[\textbf{{Z}}])^T]=\textbf{{H}} \mathsmaller{\sum\nolimits_\textbf{{x}}} \textbf{{H}}^T + \sigma^2 \textbf{I}
\end{equation}

Now the task is to find a basis matrix of $\mathcal{C}(\textbf{{H}} \mathsmaller{\sum\nolimits_\textbf{{x}}} \textbf{{H}}^T)$. First, we perform a singular value decomposition (SVD) of $\Sigma_\textbf{{Z}}$ and obtain unitary matrix $\textbf{U}$. The rank of
$\textbf{{H}} \mathsmaller{\sum\nolimits_\textbf{{x}}} \textbf{{H}}^T$ is $n$. So, $n$ columns of $\textbf{U}$ form a basis of $\mathcal{C}(\textbf{{H}} \mathsmaller{\sum\nolimits_\textbf{{y}}} \textbf{{H}}^T)$. As $\mathcal{C}(\textbf{{H}} \mathsmaller{\sum\nolimits_\textbf{{x}}} \textbf{{H}}^T)$ is equivalent of $\mathcal{C}(\textbf{{H}})$~\cite{Kim6996007}, the same $n$ columns of $\textbf{U}$ form a basis of $\mathcal{C}(\textbf{{H}})$. As a result, attackers can construct attacks using the subspace information described above without knowing the information of the original $\textbf{H}$ matrix. More information and proof regarding this type of measurement subspace based attack strategy can be obtained from~\cite{Kim6996007}.

PCA can also be used to design such blind attack as demonstrated in~\cite{yu7001709}. PCA is a multivariate statistical technique widely used for dimensionality reduction and data transformation. It can transform the correlated observations into uncorrelated variables which are known as principal components. These orthogonal principal components are the linear combinations of the original observations. After a successful PCA transformation, we obtain the vector of principal components $\textbf{u}$ and the transformation matrix $\tilde{\bf{M}}$~\cite{yu7001709}. Therefore, the PCA relationship can be represented as:
\begin{equation}\label{}
\tilde{\bf{M}}^T \textbf{Z}= \textbf{u}
\end{equation}
Now, measurement matrix $\textbf{Z}$ can be approximated by,
\begin{equation}\label{pcaalllbl2}
\textbf{Z} \approx  \begin{bmatrix}
{\tilde{\bf{M}}_{1,1}} & {\tilde{\bf{M}}_{1,2}} & \cdots &  {\tilde{\bf{M}}_{1,m}} \\
\vdots & \vdots & \ddots & \vdots  \\
{\tilde{\bf{M}}_{n,1}} & {\tilde{\bf{M}}_{n,2}} & \cdots & {\tilde{\bf{M}}_{n,m}} \\
\vdots & \vdots & \ddots & \vdots  \\
{\tilde{\bf{M}}_{m,1}} & {\tilde{\bf{M}}_{m,2}} & \cdots &  {\tilde{\bf{M}}_{m,m}} \\
\end{bmatrix}\left[
   \begin{array}{c}
\textbf{u}_1\\
 \vdots\\
\textbf{u}_n\\
 \vdots\\
\textbf{u}_m\\
  \end{array}
  \right]
  \end{equation}
where the principal components and corresponding eigenvectors are arranged based on their eigenvalues in descending order.
As the rank of the original $\textbf{H}$ matrix is $n$, authors in~\cite{yu7001709} suggest to consider only \textit{n} principal components. Hence (\ref{pcaalllbl2}) can be rewritten as follows:
\begin{equation}\label{}
\textbf{Z} \approx  \begin{bmatrix}
{\tilde{\bf{M}}_{1,1}} & {\tilde{\bf{M}}_{1,2}} & \cdots &  {\tilde{\bf{M}}_{1,n}} \\
\vdots & \vdots & \ddots & \vdots  \\
{\tilde{\bf{M}}_{m,1}} & {\tilde{\bf{M}}_{m,2}} & \cdots & {\tilde{\bf{M}}_{m,n}} \\
\end{bmatrix}\left[
   \begin{array}{c}
\textbf{u}_1\\
 \vdots\\
\textbf{u}_n\\
  \end{array}
  \right]
  \end{equation}
\begin{equation}\label{halllbl}
  \equiv \textbf{H}_{pca} \textbf{u}_{pca}
\end{equation}
where the reduced transformation matrix has been considered as $\textbf{H}_{pca}$, which is then used for stealthy attack construction. In that case, the attack vector will be,
\begin{equation}\label{avectorlbl}
  \textbf{a}_{pca}= \textbf{H}_{pca}\textbf{c}
\end{equation}
where $\textbf{c}$ is an arbitrary non-zero vector of length $n$~\cite{yu7001709}.
Therefore, the attacked measurement vector becomes
\begin{equation}\label{attackvecnownewlbl}
\textbf{z}_{attack} = \textbf{z}  +\textbf{a}_{pca}
\end{equation}
The proof of such stealthy attack strategy is given in~\cite{yu7001709}.

\subsection{Approximation of the Rank of the System Jacobian}\label{ssslbl}
\begin{figure}[h]
  \centering
    \includegraphics[width=0.5\textwidth]{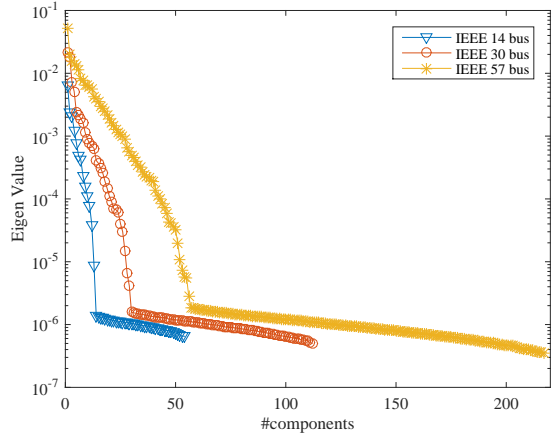}\\
  \caption{Eigenvalue analysis of different sensor measurements for three benchmark test systems}\label{eigvalAll3testsystemslbl}
\end{figure}

During the PCA based attack construction, authors in~\cite{yu7001709} suggest to consider only \textit{n} number of principal components where \textit{n} is chosen based on the rank of the system Jacobian $\textbf{H}$ matrix. Similar consideration is observed in~\cite{Kim6996007}.
As the rank of the Jacobian matrix is equal to the number of system states~\cite{yu7001709}, attacker can only get the information of the total number of system states if the system Jacobian $\textbf{H}$ is known. If it is assumed that the attacker has no knowledge on the $\textbf{H}$ matrix, it is also logical to assume that the rank of the $\textbf{H}$ matrix, i.e., $n$ is also unknown to the attacker.
In this section, we describe a heuristic based on eigenvalue analysis to approximate the rank of $\textbf{H}$ matrix and use it to construct a stealthy attack in Section 5 when $n$ (number of system states) is not known. In Fig.~\ref{eigvalAll3testsystemslbl}, we plot the eigenvalues of the principal components obtained from the measurement matrix using PCA for three IEEE benchmark test systems. Based on the eigenvalues, the contribution of each component can be estimated in order to make an approximation of the measurement subspace. In Fig.~\ref{eigvalAll3testsystemslbl}, we observe that the first few components have larger values compared to the remaining components.  We refer to these principal components as the \textit{influential} principal components and the total number of influential principal components are denoted by $\rho$. The dimension of the new projected space is reduced by considering only $\rho$ principal components and is calculated as follows:
\IncMargin{.4em}
\begin{algorithm}[!h]
\SetAlgoLined
\SetKwInOut{Input}{input}
\Input{$\textbf{v}$, i.e., $\textbf{v}={v_1,..,v_m}$ is the vector of eigenvalues of $\textit{m}$ principal components}
\ initialize $\prod=\sum_{i=1}^m v_i$, $\kappa=0$\ and counter $\textit{$\rho$}$;
\While{$\kappa \leq \Gamma$}{$\kappa=\kappa+\textbf{v}(\rho)$\\
increment $\rho$}
\KwResult{Output $\rho$}
\caption{Calculation of the number of influential principal components ($\rho$)}
\label{cumsumlbl}
\end{algorithm}\DecMargin{0.4em}

In Algorithm~\ref{cumsumlbl}, the eigenvalues are arranged in descending order in vector $\textbf{v}$ and
$\rho$ is the number of influential principal components which is obtained using a precision threshold $\Gamma$ by identifying the knee of the scree plot~\cite{Valenzuela6362259} shown in Fig.~\ref{eigvalAll3testsystemslbl}.
The precision threshold $\Gamma$ determines the size of regular subspace and approximates the noiseless measurement matrix.
Here, $\Gamma$ is a tunable parameter which is obtained empirically using trial and error until a high successful stealthy attack construction rate is achieved.
We observe that the precision threshold value 0.995 leads to high successful rate of stealthy attack construction (shown in Section~\ref{RnD}).
The  $\Gamma=0.995$ indicates that we consider 99.5\% of the noisy measurement matrix to approximate the regular subspace.
As the noise vector is independent and identically distributed (i.i.d), the above eigenvalue based dimension selection approach will also suppress the noise. We observe that the attacker can construct the attack based on the dimension ($\rho$), obtained using the eigenvalue analysis explained above, which shows almost the same stealthy characteristics as for the case with known $n$. The experimental validation is demonstrated in Section~\ref{RnD}.\\

\section{Attack in the presence of grossly corrupted measurements}\label{pmethod1sec}
In Section 4.2 and 4.1, we showed how to solve the problem of unknown system states and system Jacobian information. In this section we show how to construct a stealthy FDI attack in the presence of gross measurement errors in the data.

The data-driven attack described in the above sections is completely dependent on the measurement data. In an industrial smart grid, corrupted measurement is fairly common in sensor data, due to device malfunction and communication error. It is well established in theory that PCA cannot handle grossly corrupted data~\cite{Candes:2011:RPC:1970392.1970395,Lin2009aug}. Furthermore, we demonstrate through experiments (see Section 6.3) that the previously mentioned strategy (in Section 4.1~\cite{Kim6996007,yu7001709}) does not work in the presence of gross errors. In this section, we show how to construct a stealthy FDI attack even in the presence of gross errors. Typically, grossly corrupted measurements are only a small fraction of the total number of measurements. Therefore, gross error matrix can be considered as a sparse matrix. On the other hand, slowly varying system states lead to a low-rank measurement matrix. Hence we utilize a low-rank and sparse matrix separation technique to estimate the original low-rank measurement matrix by separating the gross errors. Next, the estimated measurement matrix is used for attack construction. We formulate the problem as follows.

The original measurement matrix is a low rank matrix~\cite{Liu6740901} and the gross errors can be assumed to be sparse. Therefore, sparse optimization technique can be used to approximate the original low-rank measurement matrix from the observed measurement matrix (with gross errors). If the original low-rank measurement matrix is $\textbf{A}$, sparse matrix of missing values is $\textbf{E}$, and the observed measurement matrix with gross errors is $\textbf{Z}$, then we can form the relationship as follows:
\begin{equation}\label{}
{\textbf{Z}}={\textbf{A}}+{\textbf{E}}
\end{equation}
Now, we can formulate the problem as a \textit{matrix recovery} problem and the exact recovery of ${\textbf{A}}$ and sparse ${\textbf{E}}$ can be represented mathematically as~\cite{Anwar2016pes}:
\begin{equation}\label{rpcatheorylbleq}
min~~\| \textbf{A} \|_* + \lambda\| \textbf{E} \|_1, ~~~~s.t.~~ \textbf{Z}=\textbf{A}+\textbf{E}
\end{equation}
In this convex optimization problem, $\|.\|_*$ and $\|.\|_1$ denotes the nuclear norm and $l_1$ norm of a matrix, respectively, and $\lambda$ is a positive weighting parameter~\cite{Lin2009fast}. To solve this problem, we use augmented lagrange multiplier (ALM)~\cite{Lin2009aug} method as discussed below~\cite{Anwar2016PAISI}:

The ALM method can be used for the general constraint optimization problem as follows:
\begin{equation}\label{}
min~~f(x),~~~~s.t.~~~h(x)=0
\end{equation}
Using the ALM method, the objective function of the above optimization problem can be written as a lagrangian function:
\begin{equation}\label{}
L(x,Y,\mu)=f(x)+\langle \gamma, h(x) \rangle + \frac{\mu}{2} \|h(x)\|^2_F
\end{equation}
where $\gamma$ is the \textit{lagrange multiplier} and $\mu$ is a positive scalar.
Considering $x=(\mathbf{A},\mathbf{E}),~ f(x)=\| \mathbf{A} \|_* + \lambda\| \mathbf{E} \|_1~and~h(x)=\mathbf{Z-A-E}$ ,
the Lagrangian function can be written as:
\begin{multline}\label{}
L(\mathbf{A,E},\gamma,\mu)=f(x)=\| \mathbf{A} \|_* + \lambda\| \mathbf{E} \|_1 +
 \langle \gamma, (\mathbf{Z-A-E}) \rangle + \frac{\mu}{2} \| \mathbf{Z-A-E} \|^2_F
\end{multline}
The solution optimization process is driven by the following two update steps,
\begin{equation}\label{Auplbl}
\mathbf{A}_{k+1}= \arg~\min~L(\mathbf{A}, \mathbf{E}_k,\gamma_k,\mu_k)
\end{equation}
\begin{equation}\label{}
\mathbf{E}_{k+1}= \arg~\min~L(\mathbf{A}_{k+1}, \mathbf{E},\gamma_k,\mu_k)
\end{equation}
Eq.~(\ref{Auplbl}) can be computed from the soft-shrinkage formula~\cite{Liu6740901}, using an iterative thresholding (IT) approach that uses the singular value decomposition (SVD) of the matrix $(\mathbf{Z}-\mathbf{E}_k+\mu_k^{-1} \gamma_k)$~\cite{Lin2009aug}. After performing the SVD, the unitary matrix $\mathbf{U}$, $\mathbf{V}$ and the rectangular diagonal matrix $\mathbf{S}$ is obtained. Then, $\mathbf{A}$ is updated as,
\begin{equation}\label{}
\mathbf{A}_{k+1}=\mathbf{U} \xi_{\mu_k^{-1}}[\mathbf{S}]\mathbf{V}^T
\end{equation}
and $\mathbf{E}$ is updated as,
\begin{equation}\label{}
\mathbf{E}_{k+1}=\xi_{\lambda \mu_k^{-1}}[\mathbf{Z}-\mathbf{A}_{k+1}+ \mu_k^{-1} \lambda_k]
\end{equation}
here $\lambda=1/\sqrt{max(m,t)}$ and $\xi$ is a soft-thresholding (shrinkage) operator, defined as~\cite{Lin2009aug}:
\begin{equation}\label{}
\xi_\varepsilon[x]=\left\{ \begin{array}{ll}
 ~~x-\varepsilon,~~~if~x> \varepsilon\\
 ~~x+\varepsilon,~~~if~x< -\varepsilon\\
  ~~0,~~~otherwise,
\end{array} \right.
\end{equation}
During each of the iterations, $\gamma$ and $\mu$ are updated as follows:
\begin{equation}\label{}
\gamma_{k+1}=\gamma_k+\mu_k(\mathbf{Z}-\mathbf{A}_{k+1}-\mathbf{E}_{k+1})
\end{equation}
\begin{equation}\label{}
\mu_{k+1}=\Omega\mu_k
\end{equation}
where $\Omega$ is a positive constant. The optimization process continues until the convergence criteria is satisfied. The convergence is checked based on the relative error using (\ref{raelbl}) against a tolerance, $\tau$.
\begin{equation}\label{raelbl}
c_{k}^{idx}= \| \mathbf{Z}-\mathbf{A}_{k+1}-\mathbf{E}_{k+1}\|_F/\|\mathbf{Z}\|_F;
\end{equation}
The proof of convergence of the ALM algorithm can be obtained from~\cite{Lin2009aug}. Once the algorithm has converged, the principal component matrix, $\textbf{H}_{pca}$ of the recovered measurement matrix $\mathbf{A}$ is obtained by following the procedure discussed in Section~\ref{Hunknown}. Next, we use (\ref{avectorlbl}) to construct an FDI attack. The whole procedure of the FDI attack construction considering grossly corrupted measurements is presented in Algorithm 2.
\begin{figure}[h]
  \centering
  \includegraphics[width=0.6\textwidth]{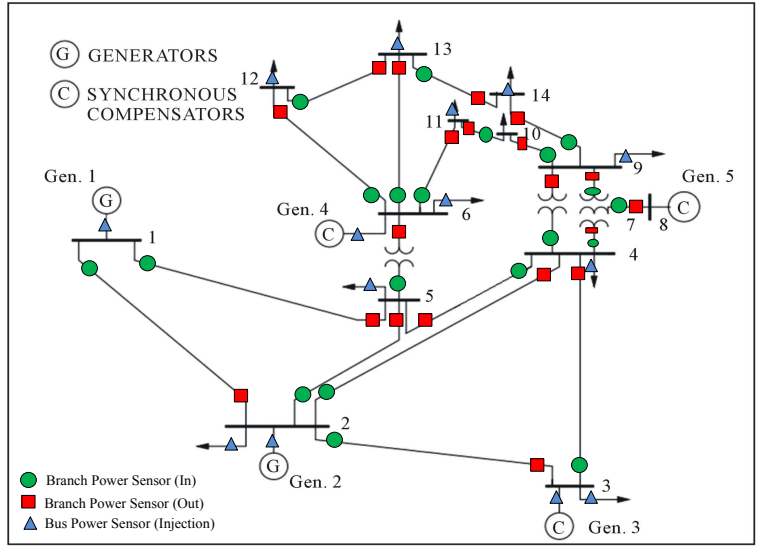}\\
  \caption{Location and types of power measurement sensors for an IEEE 14 bus test system.}\label{14buslbl}
\end{figure}

\IncMargin{.4em}
\begin{algorithm}[!h]
\SetAlgoLined
\SetKwInOut{Input}{input}
\Input{Grossly corrupted measurement matrix, $\mathbf{Z}$}
\ {[${\textbf{A}}$,~${\textbf{E}}$]=alm(\textbf{Z}) // use ALM based convex optimization to generate true measurement matrix $\mathbf{A}$ by separating sparse gross error matrix $\mathbf{E}$}  \;
\ [$\mathbf{\tilde{M}}$,\textbf{v}]=pca(\textbf{Z})~~// calculate $\mathbf{\tilde{M}}$ and $\textbf{v}$ using principal component analysis  \;
\ Calculate $\rho$ using the eigenvalue analysis following Algorithm 1\;
\ Obtain $\textbf{H}_{pca}$ using $\rho$ principal components\;
\ Generate a non-zero vector $\textbf{c}$ of the length $\rho$ and construct attack vector, $\textbf{a}_{pca}=\textbf{H}_{pca}\textbf{c}$ \;
\ Calculate manipulated sensor measurements, $\textbf{z}_{attack}=\textbf{z}+ \textbf{a}_{pca}$ \;
\caption{ALM based FDI attack strategy}
\end{algorithm}\DecMargin{0.4em}

\textbf{Detailed Explanation of Algorithm 2:}
The detailed description of Algorithm 2 is analysed step by step below:\\

\textbf{Step 1}: In this step, the attacker uses ALM based sparse matrix optimization technique~\cite{Lin2009aug} to separate the gross errors (\textbf{E}) from the measurement matrix (\textbf{Z}) in order to obtain the gross-error free matrix (\textbf{A}).\\

\textbf{Step 2}: PCA transformation is then performed on the gross-error free matrix (\textbf{A}) to obtain a new projected subspace of the measurements.
Here, PCA provides the transformation matrix $\mathbf{\tilde{M}}$ (that contains all the $m$ principal components), and corresponding eigenvalues in vector $\textbf{v}$, where $\textbf{v}={v_1,..,v_m}$. In $\mathbf{\tilde{M}}$, the first column is the most informative principal component and the significance of principal components decreases gradually as the column number increases. \\

\textbf{Step 3}: Using the eigenvalues from Step 2, and applying the heuristic described in Algorithm 1, we obtain $\rho$ which is used in next to generate stealthy attack. \\

\textbf{Step 4}: In order to generate an attack, we first obtain reduced transformation matrix $\mathbf{H}_{PCA}$ from $\mathbf{\tilde{M}}$ by taking the first $\rho$ principal components using the approach discussed in Eq.~(\ref{pcaalllbl2})-(\ref{halllbl}). \\

\textbf{Step 5}: We construct the stealthy attack vector using $\textbf{a}_{pca}=\textbf{H}_{pca}\textbf{c}$ of Eq.~(\ref{avectorlbl}), where $\textbf{c}$~\cite{Liu:2011:FDI:1952982.1952995} is a non-zero Gaussian random vector of length $\rho$.\\

\textbf{Step 6}: The injected attack vector ($\textbf{a}_{pca}$) modifies the original measurements ($\textbf{z}$) as, $\textbf{z}_{attack}=\textbf{z}+ \textbf{a}_{pca}$ following Eq.~(\ref{attackvecnownewlbl}).    \\

\section{Experimental Evaluation:}\label{RnD}
\subsection{Test Systems:}
All experiments were conducted using benchmark IEEE power systems, which include IEEE 14 bus, IEEE 30 bus and IEEE 57 bus test systems. Power system data can be obtained from~\cite{wtestcases,Zimmerman5491276}.
There is no known publicly available real-world smart grid cyber-attack data~\cite{Valenzuela6362259}. Hence, realistic power system simulation is carried out using Matlab based simulation tool MATPOWER~\cite{Zimmerman5491276}. MATPOWER is widely used for simulating power system data and reflects a realistic simulation environment for the real-life complex power systems~\cite{Zimmerman5491276,Esmalifalak6102326,yu7001709}. The state estimation formulation and attack construction strategies are also implemented in Matlab on a PC with an Intel(R) core i7 @ 3.4 GHz- 3.4 GHz processor and 16 GB of RAM.
\begin{figure*}[hb]
  \centering
  \includegraphics[width=0.7\textwidth]{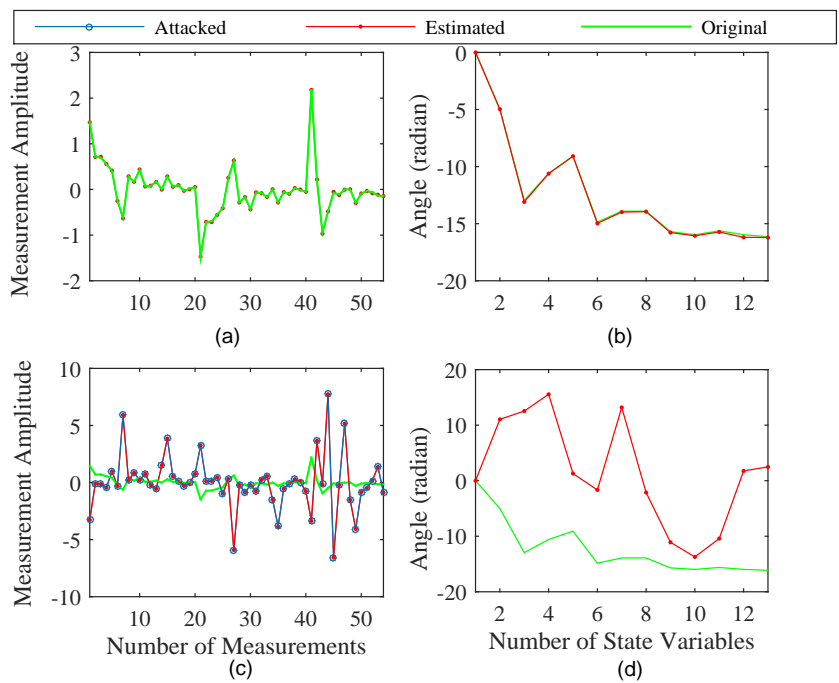}\\
  \caption{Sensor measurements and State variables are plotted for the \textit{no~attack~case} in (a)-(b) and the \textit{attack~case} with known system information in (c) and (d), respectively.}\label{FDItheoryhelp1lbl}
\end{figure*}
\subsection{\textbf{Comparison of different FDI attack strategies without gross errors} ({Contribution 1}):}
In this section, we evaluate the performance of different attack construction strategies and compare their stealthiness against the case when no attack vector is injected. First, we consider a single scenario to illustrate the attack strategies and then we consider a monte carlo simulation of 1000 runs to evaluate the performance over a broad selection of scenarios and test setups. Here, we consider the voltage angles as system states, which is obtained after a successful SE. The vector of measurement signals ($\textbf{z}$) is necessary for SE operation, which is obtained from the \textit{power flow} solution of the test system using MATPOWER. Now, data-driven \textit{blind} FDI attack needs multiple observations. Hence, $100$ samples (observations) are generated
using the same mean of ($\textbf{z}$) with gaussian distribution similar to~\cite{Kim6996007}. The impact of measurement noise is also considered by introducing zero mean Gaussian noise with the ideal power flow measurements. We have considered signal-to-noise ratio (SNR) between $20\sim35$ db during the simulation. All the branch power flows (both incoming and outgoing) and bus power injections have been considered to generate the measurement vector. For example, the IEEE 14 bus test system has 20 line sections (branches) and 14 buses (nodes). Therefore, the total measurements consist of 20 incoming power flow sensors, 20 outgoing power flow sensors, and 14 power injection sensors (total 54 sensors). These sensors are shown in Fig.~\ref{14buslbl}, where the sensors are marked with individual symbols. Similarly, the IEEE 30 bus test system has 41 line sections and 30 buses which includes 112 measurement sensors. The IEEE 57 bus test system has 80 line sections and 57 buses, which includes a total of 217 measurement sensors. The IEEE 14 bus, 30 bus and 57 bus systems have 13, 29 and 56 unknown system states, respectively. Note, the number of unknown system states is always at least one less than the total number of buses as one bus is considered as a reference bus during power flow simulation and removed from the unknown system state vector.

\textit{\textbf{Attack construction using known system information:}}
In our test setup, the IEEE 14 bus system has 54 measurements and 13 states, which provides a degree of freedom of 41.
Following a chi-square test considering 95\% confidence interval, the anomaly threshold for BDD module becomes 56.94~\cite{abur2004power}. Under normal operating conditions (no attack scenario), the estimated system states follow the true states. One such scenario is shown in Fig.~\ref{FDItheoryhelp1lbl}(a)-(b) which has a residual value of 39.88 (well below the threshold) obtained using the IEEE 14 bus test system. Note that, Fig.~\ref{FDItheoryhelp1lbl}(a) and Fig.~\ref{FDItheoryhelp1lbl}(b) both contains two graphs each (one for normal and the other for estimated), but due to superimposition the two graphs cannot be distinguished (residual = 39.88). We simulate the attack strategy proposed by Liu et al. with known system information (system Jacobian \textbf{H}) in Fig.~\ref{FDItheoryhelp1lbl} (c)-(d). From Fig.~\ref{FDItheoryhelp1lbl} (d), the estimated states are too far away from the original system states and the SE module minimizes the cost assuming the attacked signal is the original signal (see Fig.~\ref{FDItheoryhelp1lbl}(c)). Interestingly, for the attacked case, we obtain the same residual value (39.88) although the system states have been changed heavily. Thus, the attack remains hidden as the residual of the attacked case is also below the threshold of the existing detection technique. Therefore, a successful attack can be constructed using known system information.
\begin{figure*}[t!]
    \centering
    \begin{subfigure}[t]{0.5\textwidth}
        \centering
        \includegraphics[height=2.4in]{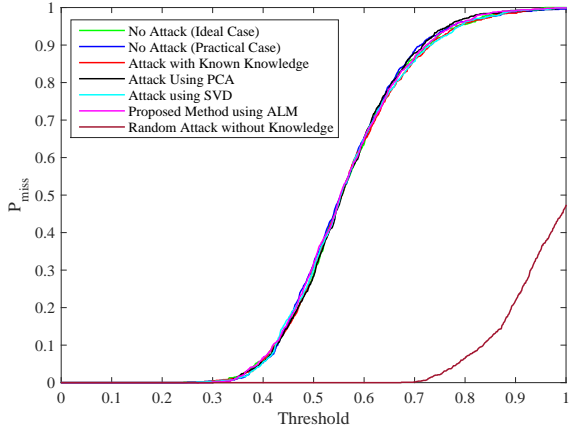}
        \caption{IEEE 14 bus test system}
    \end{subfigure}%
    \begin{subfigure}[t]{0.5\textwidth}
        \centering
        \includegraphics[height=2.4in]{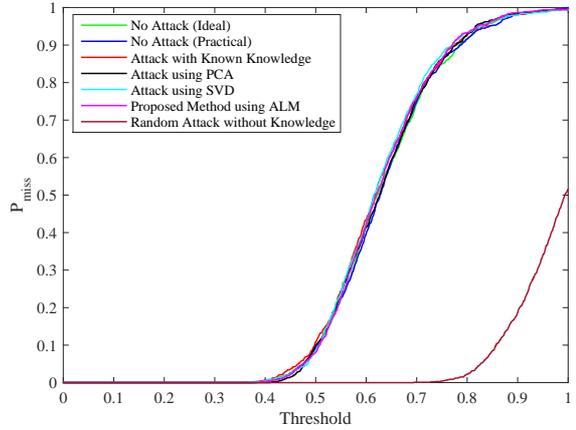}
        \caption{IEEE 30 bus test system}
    \end{subfigure}
    \begin{subfigure}[t]{0.5\textwidth}
        \centering
        \includegraphics[height=2.4in]{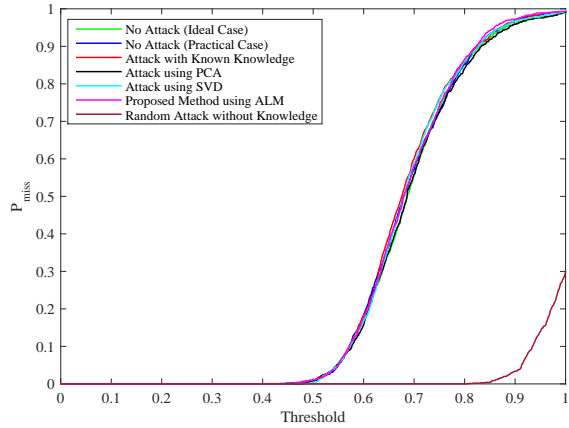}
        \caption{IEEE 57 bus test system}
    \end{subfigure}
    \caption{Probability of misdetection versus the detection threshold for different attack strategies.}
    \label{pmisslbl4threefigs}
\end{figure*}

\textit{\textbf{Attack construction without system information:}}
If the system information (system Jacobian \textbf{H}) is not known, still the attack can be constructed based on the principal components of the measurement subspace. The state vectors at different time instances are independent and identically distributed (i.i.d) which follows a Gaussian distribution with mean equal to the operating point defined by the test system\cite{Kim6996007}. Here, 500 time instances were considered during the creation of the measurement matrix following the relationship expressed in Eqn.~\ref{dcselbl}. Measurement noise is considered to be between $20\sim35$ dB.
The data-driven FDI attack without power system knowledge is constructed using the method discussed in Section~\ref{Hunknown}. Fig.~\ref{pmisslbl4threefigs} shows the probability of false negative, $P_{mis}$ (attack incorrectly identified as normal) under different attack scenarios considering the IEEE 14 bus, 30 bus and 57 bus test systems, respectively. For each of the three figures, we plot the following attack strategies- (1) FDI attack based on system knowledge \textbf{H} as proposed by Liu et al~\cite{Liu:2011:FDI:1952982.1952995}, (2) blind FDI attack using PCA~\cite{yu7001709}, (3) data-driven FDI attack using SVD~\cite{Kim6996007}, (4) random attack without system knowledge and (5) ALM based proposed attack model.
\begin{figure*}
  \centering
  \includegraphics[width=0.67\textwidth]{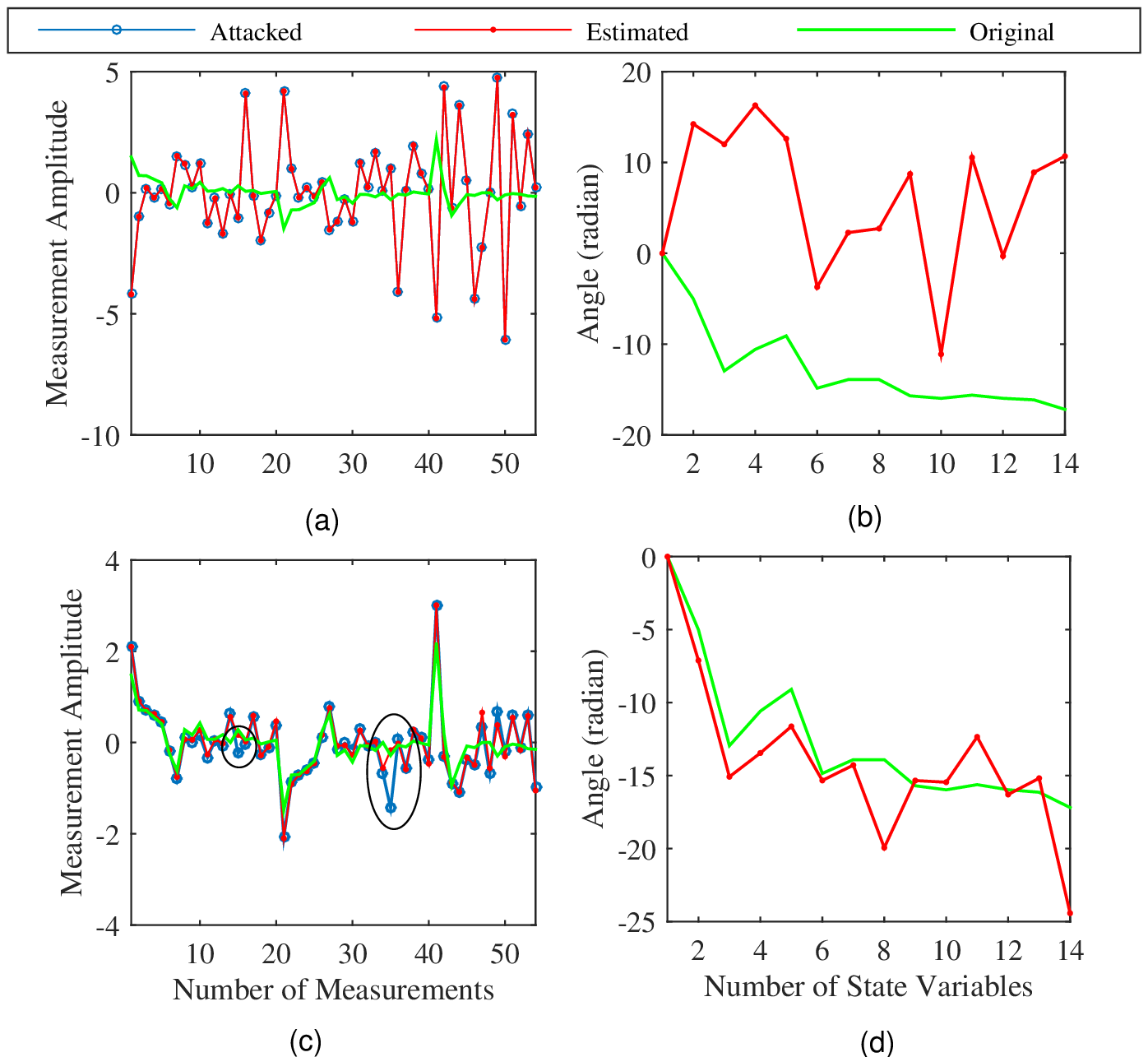}\\
  \caption{Effect of PCA based FDI attack construction strategy on sensor measurements and estimated state variables, without gross error~(a)-(b) and with gross error~(c)-(d), respectively.}
  \label{FDItheoryhelp2}
\end{figure*}

In Fig.~\ref{pmisslbl4threefigs}(a), first we plot the probability of detection for a normal scenario, referred to as the \textit{no attack} case, for the IEEE 14 bus test system considering both noiseless measurements (ideal) and measurements with Gaussian noises (practical). Next, we simulate each of the above discussed attack strategies 1000 times considering different noise vectors and plot the corresponding $P_{mis}$ for different threshold values. We observe that all attack strategies have almost the same stealthy nature as the \textit{no attack} case except the \textit{random attack} strategy which performs very poorly. For any specific threshold value, the \textit{random attack} without system knowledge has the lowest $P_{mis}$. Therefore, it can be easily distinguishable from the \textit{no attack} case. Other than the \textit{random attack}, all other attack strategies including our proposed method conform with the \textit{no attack} case and become indistinguishable in the existing BDD module. Similar observations were made when performing experiments on the two other benchmark test systems which are plotted in Fig.~\ref{pmisslbl4threefigs}(b) and Fig.~\ref{pmisslbl4threefigs}(c).

\subsection{\textbf{Inefficiencies of existing attack strategies in the presence of gross errors} ({Contribution 2}):}\label{grosserrorlbl}
In the previous section, we have shown that all the attack strategies (except the random attack) maintain stealthy characteristics similar to that of the \textit{no attack} case, and remain hidden, however, this observation is only true when the measurement matrix does not have a missing value or grossly corrupted measurement. We observe that, a stealthy FDI attack based on the PCA method becomes detectable by the State Estimator detection modules if the measurement data contains one or more grossly corrupted measurement data or missing value. This limitation of traditional PCA is discussed in~\cite{Candes:2011:RPC:1970392.1970395}. Here, we show an example of the PCA based attack using sensor measurements that contain only one single gross error. We consider the IEEE 14 bus test system and create a measurement matrix based on the operating conditions provided in the test system using MATPOWER. The measurement matrix contains the measurements of 54 sensors for 500 different observations. Next, we consider a single gross error in the measurement matrix, which can be any sensor value of any observation. We construct a PCA based blind FDI attack vector using the measurement matrix. This process is repeated 100 times to generate different attack cases. All of the 100 cases are detected as the residuals produced under this attack scenario are significantly larger than the detection threshold of the BDD module. One such scenario is shown in Fig.~\ref{FDItheoryhelp2}, where the measurements and state variables are plotted in (c) and (d) respectively. We see that the estimated state variables have different values than the true variables. However, the estimated measurements can not follow the attacked measurements (which is assumed to be true at the utility's detection module), which leads to a high estimation error (residual). For example, for this specific case, the Weighted Sum of Squared Error (WSSE) is $5.33*10^3$ and the detection threshold is 56.94. Therefore, the attack is easily detectable using the existing BDD module. If the measurement matrix does not include that gross error, we obtain the plots of Fig.~\ref{FDItheoryhelp2} (a) and (b) for the same experimental setup. From those figures, we see that the estimated state variables are far away from the true states but the estimated measurements coincide with the attack measurements. Therefore, the WSSE is 45.54 (clearly below the threshold) and the attack is successful. In summary, we conclude that any single gross error can make the PCA based data-driven attack detectable.
\begin{figure*}
  \centering
   \includegraphics[width=1\textwidth]{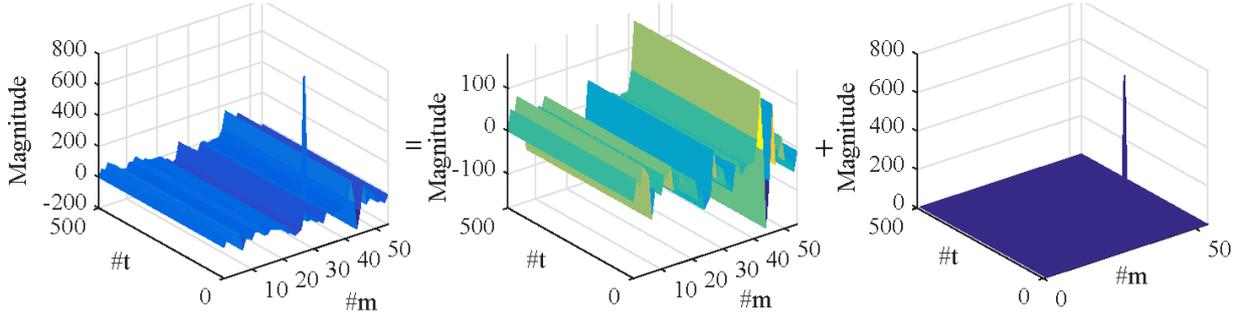}\\
  \caption{The ALM method approximates the true measurement matrix by separating the gross errors. The left plot shows the corrupted measurements with gross errors and the remaining two are its decomposition into a low-rank and sparse matrix, respectively.}\label{FDItheoryhelp3lbl}
\end{figure*}

\subsection{\textbf{Proposed attack strategy in the presence of gross errors} ({Contribution 3}):}

\begin{figure*}[hb]
  \centering
   \includegraphics[width=0.56\textwidth]{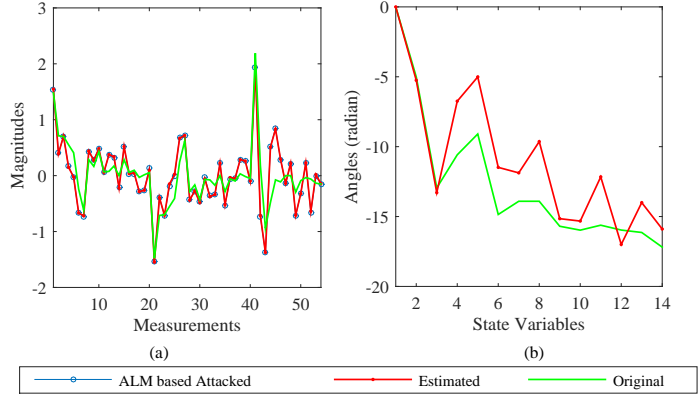}\\
  \caption{Data-driven attack construction using the ALM method considering gross error.}
  \label{FDItheoryhelp4lbl}
\end{figure*}

The poor performance of the PCA-based attack is due to its brittleness in the presence of gross error. Here, we demonstrate how to construct an undetectable FDI attack by recovering the true low-dimensional subspace of the measurement data from the noisy data. Here, we use ALM based sparse optimization technique to approximate the measurement matrix that is very close to the true matrix by separating the sparse gross error. The same experiment discussed in the previous section is repeated using the ALM based method. We observe that the ALM based sparse optimization technique can approximate the true measurement and sparse gross error very accurately, as plotted in Fig.~\ref{FDItheoryhelp3lbl}. Hence, a data-driven attack is possible based on the low-rank approximation which is very close to the true measurement matrix. In this section, we set up an experiment where we generate a measurement matrix using 500 observations for the 54 measurement sensors of the IEEE 14 bus test system in a similar manner to that discussed in the previous section. Next we add $1\%$ gross errors and missing values to the measurement matrix to create a grossly corrupted observation matrix. Here, we define gross errors as a sparse matrix with the same dimensions of the measurement matrix but a high value, which is significantly larger than the true measurements. Once the grossly corrupted measurement is generated, the aim of this work is to test whether the cyber-attacker can construct an undetectable attack. Hence, we employ the ALM based method and see that undetectable attack construction is indeed possible as the ALM method recovers a highly accurate approximation (relative error is around $10^{-8}$) of the true measurement matrix. Here, relative error is defined as:
\begin{equation}\label{relbl2}
  R_E= \frac{\|(\textbf{Z}-{\textbf{A}}-{\textbf{E}})\|_2}{\|\textbf{Z}\|_2}
\end{equation}
where, Z is the corrupted measurement, ${A}$ and ${E}$ are the approximation of the true low-rank and sparse matrix, respectively. Fig.~\ref{FDItheoryhelp3lbl} shows the state variables and the measurement vectors under the normal and ALM based attack cases. We observe that the estimated state variables possess a different value than the true states (Fig.~\ref{FDItheoryhelp4lbl}.[a]) and the estimated measurement follows the manipulated attack measurements perfectly. For this specific case, the WSSE is $32.84$ which is well below the threshold. As demonstrated, the ALM based data-driven FDI attack is successful and remains stealthy.

\subsection{\textbf{Comparison of the proposed method with benchmark sparse matrix separation techniques} ({Contribution 4}):}
In this section, we compare the performance of the proposed attack strategy in terms of accuracy and the time-efficiency with a number of sparse optimization techniques (SVT, APG and DUAL methods) considering three IEEE benchmark test systems (IEEE 14, 30 and 57 bus systems).

\begin{figure*}[t!]
    \centering
    \begin{subfigure}[t]{0.5\textwidth}
        \centering
        \includegraphics[height=2.4in]{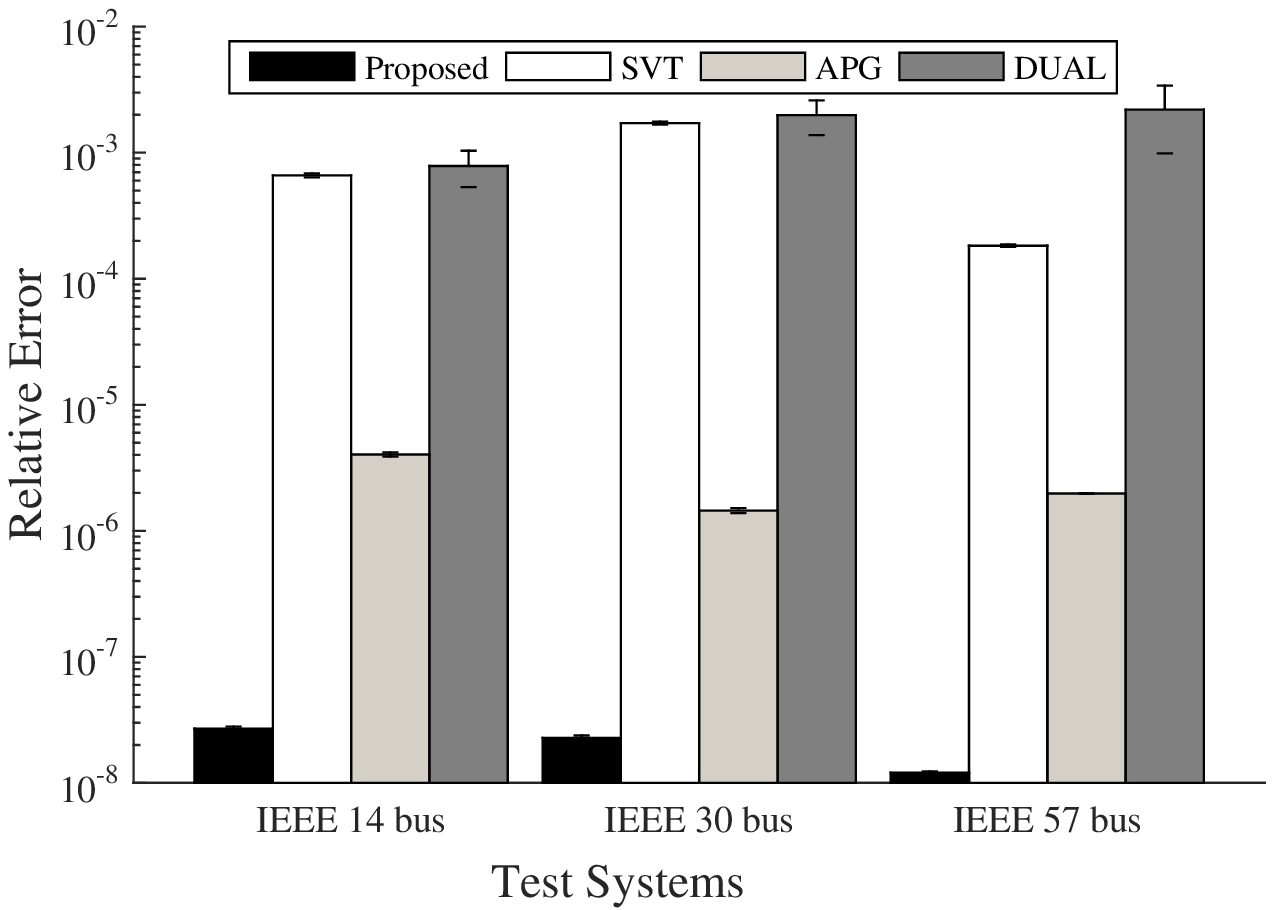}
        \caption{1\% gross error}
    \end{subfigure}%
    \begin{subfigure}[t]{0.5\textwidth}
        \centering
        \includegraphics[height=2.4in]{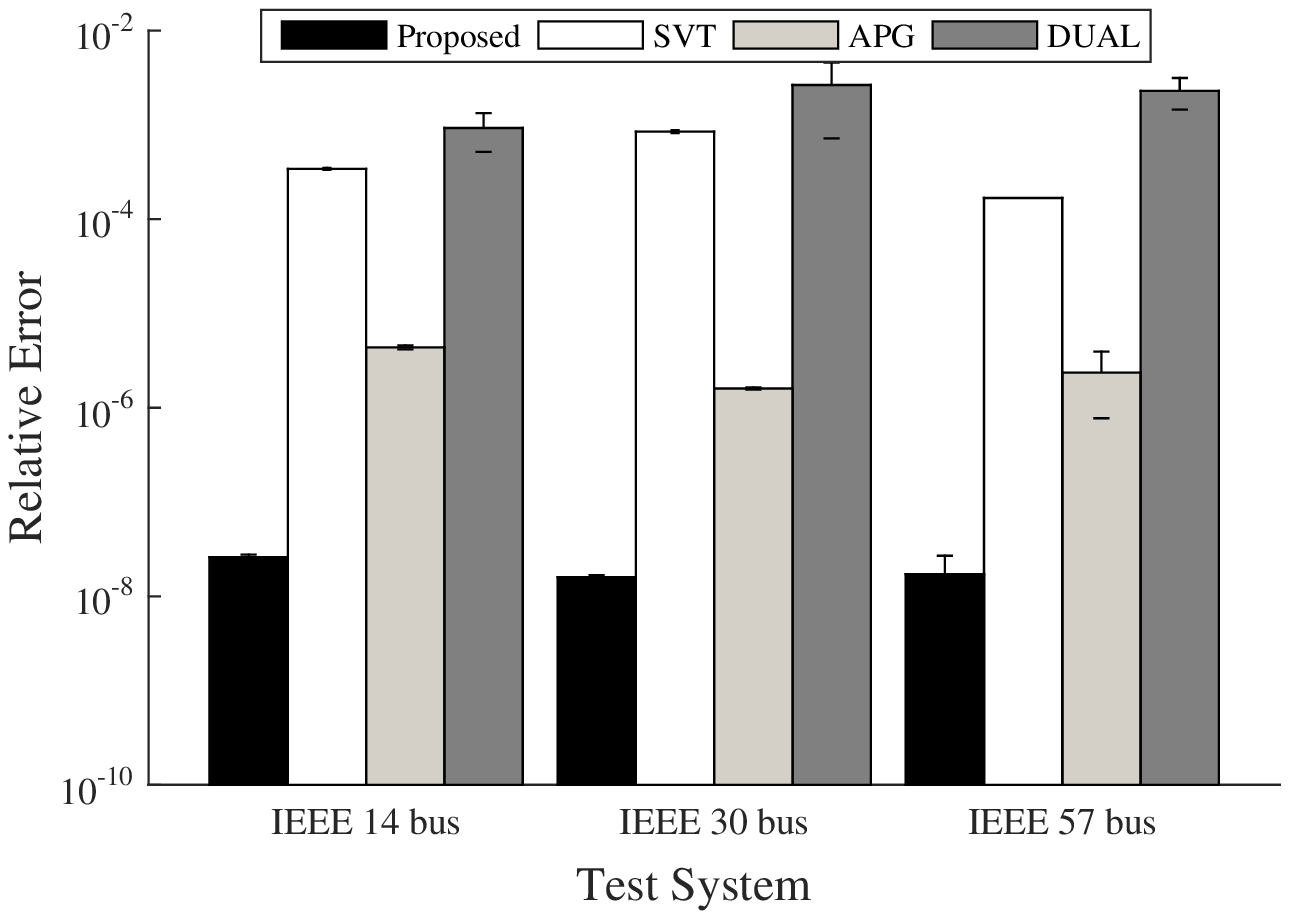}
        \caption{5\% gross error}
    \end{subfigure}
    \caption{Accuracy of approximating measurement matrix using Relative error ($R_E$) of the proposed method compared with different sparse optimization techniques (APG, SVT and Dual methods).}
     \label{accuracyallplotlbl}
\end{figure*}
\textit{\textbf{(1) The accuracy of the approximate measurement matrix (Fig.~\ref{accuracyallplotlbl})}:} The accuracy of the approximate matrix can be determined using the \textit{relative error} in Eqn.~(\ref{relbl2}). In this section we measure the performance of the ALM method in terms of accuracy- \textit{how close the approximations are to the true subspace}. We also compare the performance against three other established algorithms.
Here, we consider the IEEE 14 bus test system. First we generate a measurement vector based on the operating point defined in the test system using MATPOWER. Next, we generate a measurement matrix by considering 500 time instances where the state vectors at different time instances are Gaussian random vectors (i.i.d) with the same mean equal to the operating condition. To include the effect of the gross errors, we add a sparse matrix that has the same dimension as the true measurement matrix with random values that are significantly higher than the true measurements. We apply Algorithm 1 to the grossly corrupted matrix to construct an undetectable attack vector using the ALM, APG, SVT and Dual methods. For each method we repeat the procedure 100 times considering different measurement realizations, Gaussian noise (SNR between 20dB $\sim$ 35dB), 1\% and 5\% sparse gross errors. For each method we report the relative approximation error. Similar experiments were performed for the IEEE 30 bus and 57 bus test systems. All results are shown in Fig.~\ref{accuracyallplotlbl}
For all test setups, we consider maximum iteration equal to 3000 for all methods and initial conditions as defined in~\cite{Lin2009aug,Lin2009fast,Cai:2010:SVT:1898437.1898451}.
From Fig.~\ref{accuracyallplotlbl}, we see that the best performing algorithm in terms of less relative error is the proposed ALM method and the second best algorithm is the APG method. The other two methods, SVT and Dual, have higher relative errors than the the ALM and APG based methods. The ALM method produces less relative error for both 1\% and 5\% sparsity as the low-dimensional approximation of the true measurements is more accurate for this algorithm than for the three other methods.

\begin{figure*}[t!]
    \centering
    \begin{subfigure}[t]{0.5\textwidth}
        \centering
        \includegraphics[height=2.25in]{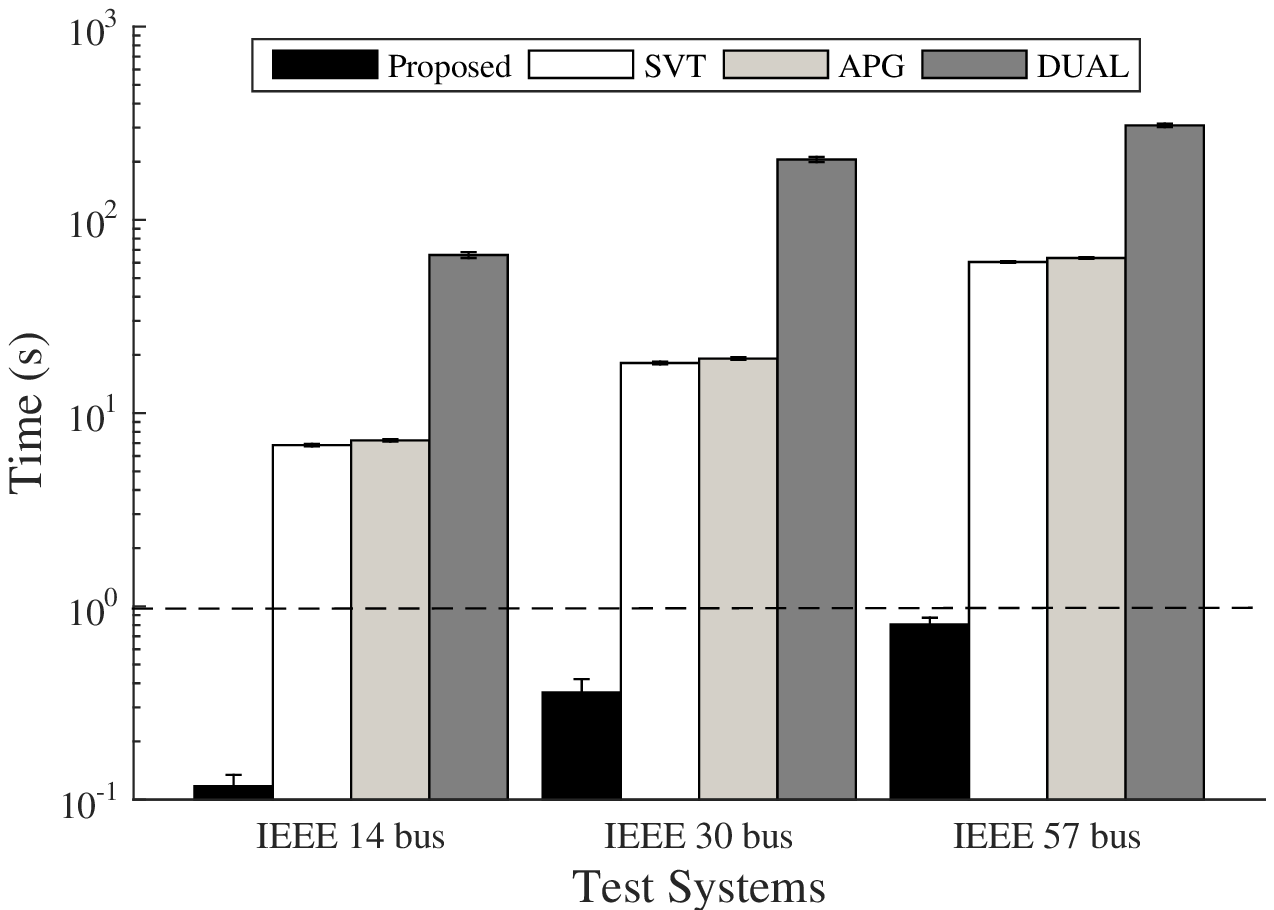}
        \caption{1\% gross error}
    \end{subfigure}%
    \begin{subfigure}[t]{0.5\textwidth}
        \centering
        \includegraphics[height=2.25in]{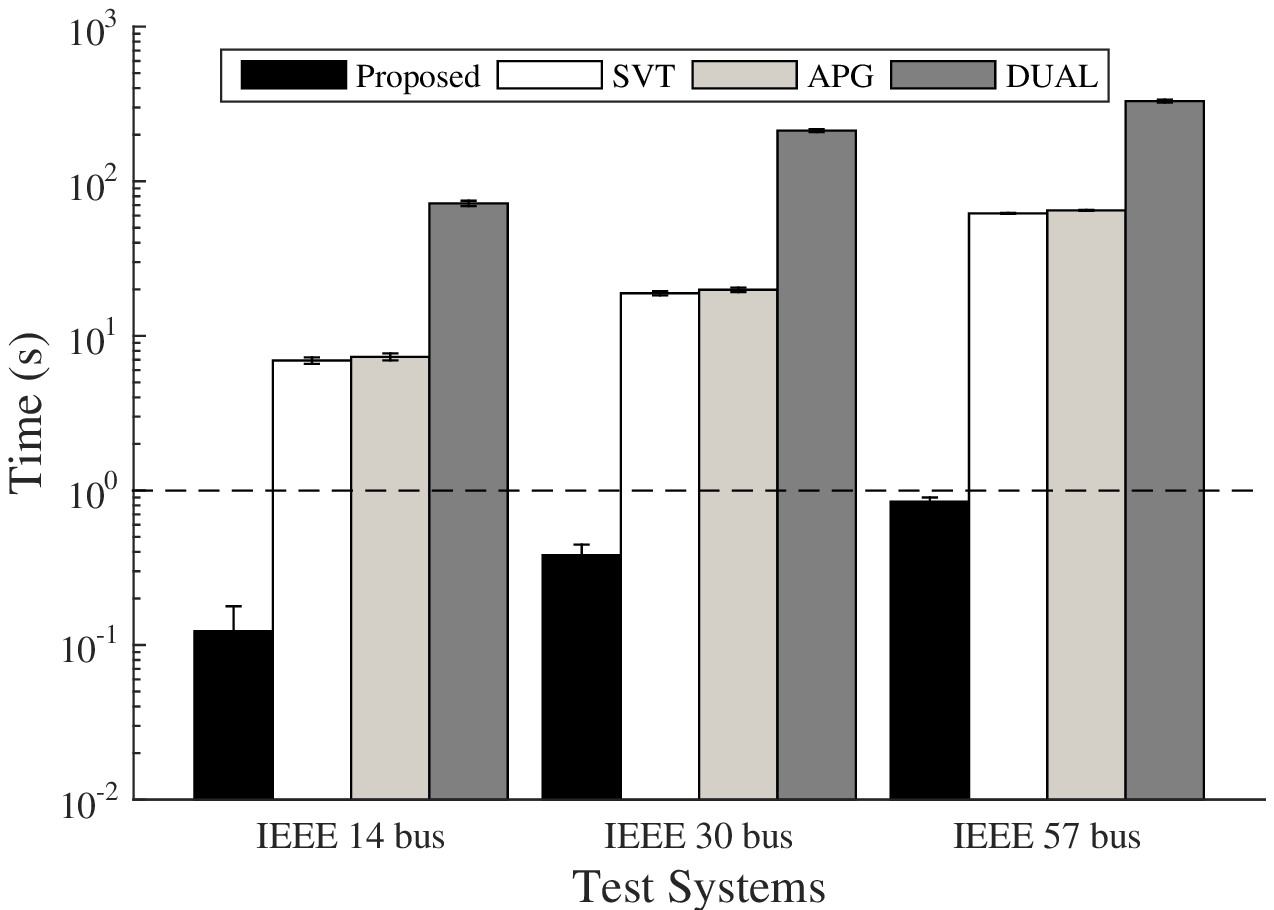}
        \caption{5\% gross error}
    \end{subfigure}
    \caption{Time-efficiency for attack construction of the proposed method compared with different sparse optimization techniques (APG, SVT and Dual methods).}
    \label{timeplotalllbl}
\end{figure*}

\textit{\textbf{(2) The time-efficiency of the attack construction (Fig.~\ref{timeplotalllbl})}:} The attack construction also needs to meet a time requirement to increase its probability of becoming stealthy. If the time required for attack construction is too high, the power system operating conditions may change and this will increase the probability of the attack being detected. Hence, in this section we measure the performance of the ALM method in terms of efficiency- \textit{how fast can it generate an attack}. We will compare the performance against three other well-performed algorithms.

The time requirement of the above discussed scenarios for all four algorithms is plotted in~Fig.~\ref{timeplotalllbl}. Although the APG based method has close accuracy to the ALM based method (as seen in the previous section), it performs poorly in terms of solution speed as evident from Fig.~\ref{timeplotalllbl}. For example, the ALM based attack construction method takes less than 1 s (as low as 0.11s for the 14 bus test case) for all three test systems. On the other hand, the APG based attack construction method takes a minimum of 7.2s (average 7.22s) for the 14 bus test system and 62.9s (average 63.5) for the 57 bus test system. Therefore, the ALM method outperforms the APG method as it has lower relative errors and faster processing capabilities, as observed for all three test systems in all test setups. The SVT and the Dual methods also require longer computational time.

\section{Conclusion}\label{SectionFinish}
The vulnerability of smart grid state estimation to an FDI attack was highlighted in this paper.
Existing attack strategies assume some system information is known, including the Jacobian matrix $\textbf{H}$ and the number of system states $(n)$ is known to the adversary. In practice, neither the information on $\textbf{H}$ nor $n$ is available to the adversary. This paper proposed a technique that can create an attack vector without knowing $\textbf{H}$ or ${n}$. We have shown that the stealthiness of the PCA based existing \textit{blind} attack cannot be guaranteed if the measurement data contains any gross errors. We have argued that an attacker can circumvent this problem by using an alternative form of attack. Here, we proposed an ALM based attack construction strategy where the original low dimensional measurement matrix (based on what the attacker can generate in a \textit{blind} attack) can be approximated by using sparse optimization. With extensive experiments that consider multiple benchmark test systems, we have demonstrated that an attacker can successfully inject ALM based hidden FDI attacks in the presence of gross errors. Different sparse optimization based attack strategies are considered to validate the proposed method. This paper concludes that the proposed ALM based FDI attack can be generated without prior system knowledge and state information and can handle different types of noise cases e.g., Gaussian errors, and grossly corrupted measurements more efficiently (in time) than existing techniques. Effective detection and prevention techniques for these types of attack are under preparation and will be presented in future work.

\section*{References}
\bibliographystyle{IEEEtran}
\bibliography{FDI_RPCA}
\end{document}